\documentclass[%
 reprint,
 onecolumn,
 longbibliography,
nofootinbib,
 amsmath,amssymb,
 aps,
prfluids
]{revtex4-2}

\usepackage{graphicx}% Include figure files
\usepackage{dcolumn}% Align table columns on decimal point
\usepackage{bm}% bold math
\usepackage{hyperref}% add hypertext capabilities
\usepackage{upgreek}
\usepackage{xcolor}
\usepackage[normalem]{ulem}
\newcommand{\stkout}[1]{\ifmmode\text{\sout{\ensuremath{#1}}}\else\sout{#1}\fi}

\newcommand{\rf}[1]{{\color{blue}#1}}

\newcommand{\rmd}{{\rm d}}

\newcommand\bb[1]{\mbox{\boldmath{$#1$}}}
\newcommand\grad{\bb{\nabla}}

\newcommand\btimes{\,\bb{\times}\,}
\providecommand\bnabla{\boldsymbol{\nabla}}

\newcommand\uv{\bb{u}}

\newcommand\bomega{\bb{\omega}}

\newcommand\Ptot{\mathcal{P}}
\newcommand\ev{\bb{e}}
\newcommand\gv{\bb{g}}
\newcommand\bSigma{\bb{\Sigma}}
\newcommand\Eu{\mathcal{E}_u}
\newcommand\Etheta{\mathcal{E}_\theta}
\newcommand\wtuv{\widetilde{\uv}}
\newcommand\wtu{\widetilde{u}}

\newcommand\wtbomega{\widetilde{\bomega}}

\newcommand\wttheta{\widetilde{\theta}}
\newcommand\wtw{\widetilde{w}}
\newcommand\wtN{\widetilde{N}}

\newcommand\wtPtot{\widetilde{\Ptot}}
\newcommand\wtEu{\widetilde{\mathcal{E}}_u}
\newcommand\wtEtheta{\widetilde{\mathcal{E}}_\theta}
\newcommand\wtbSigma{\widetilde{\bSigma}}

\newcommand{\bTomegacrossu}{\bb{\mathcal{T}}_{\omega\times u}}
\newcommand{\TN[1]}{\mathcal{T}_{N}^{(#1)}}
\newcommand\bTthetau{\bb{\mathcal{T}}_{\theta u}}
\newcommand\bTuu{\bb{\mathcal{T}}_{uu}}
\newcommand\bTvv{\bb{\mathcal{T}}_{vv}}

\newcommand\wtD{\widetilde{D}}

\usepackage{scalerel,stackengine}
\stackMath
\newcommand\reallywidehat[1]{%
\savestack{\tmpbox}{\stretchto{%
  \scaleto{%
    \scalerel*[\widthof{\ensuremath{#1}}]{\kern-.6pt\bigwedge\kern-.6pt}%
    {\rule[-\textheight/2]{1ex}{\textheight}}%WIDTH-LIMITED BIG WEDGE
  }{\textheight}% 
}{0.5ex}}%
\stackon[1pt]{#1}{\tmpbox}%
}

\newcommand\reallywidetildeB[1]{
\stackon[-8pt]{#1}{\vstretch{1.5}{\hstretch{1.8}{\widetilde{\phantom{\;\;\;\;\;\;\;}}}}}
}

\definecolor{BLUE}{RGB}{0, 0, 255}

\begin{document}

\preprint{APS/123-QED}

\title{Characterization of local energy transfer in large-scale intermittent stratified turbulent flows via coarse-graining}

%: filtering the Boussinesq equations}% Force line breaks with \\

\author{Raffaello Foldes}
 \email{raffaello.foldes@ec-lyon.fr}
\affiliation{%
  CNRS, \'Ecole Centrale de Lyon,  INSA de Lyon, Universit\'e Claude Bernard Lyon 1, Laboratoire de M\'ecanique des Fluides et d’Acoustique, F-69134 \'Ecully, France.}%
\affiliation{%
Dipartimento di Scienze Fisiche e Chimiche, Universit\`a dell'Aquila, L'Aquila, Italy
}%

\author{Raffaele Marino}
\affiliation{%
CNRS, \'Ecole Centrale de Lyon,  INSA de Lyon, Universit\'e Claude Bernard Lyon 1, Laboratoire de M\'ecanique des Fluides et d’Acoustique, F-69134 \'Ecully, France.
}%

% \collaboration{MUSO Collaboration}%\noaffiliation

\author{Silvio Sergio Cerri}
\affiliation{
 Universit\'e C\^{o}te d'Azur, Observatoire de la C\^{o}te d'Azur, CNRS, Laboratoire Lagrange, Bd de l'Observatoire, CS 34229, 06304 Nice cedex 4, France
}%

\author{Enrico Camporeale}
% \email[]{enrico.camporeale@noaa.gov}
%\thanks{}
\affiliation{School of Physical and Chemical Sciences, Queen Mary University of London, London E1 4NS, UK}
\affiliation{Space Weather Technology, Research and Education Center (SWx-TREC), University of Colorado, Boulder, Colorado 80309, USA}

\date{\today}% It is always \today, today,

\begin{abstract}
Recent studies based on simulations of the Boussinesq equations indicate that stratified turbulent flows can develop large-scale intermittency in the velocity and temperature fields, as detected in the atmosphere and oceans. In particular, emerging powerful vertical drafts were found to generate local turbulence, proving necessary for stratified flows to dissipate the energy as efficiently as homogeneous isotropic turbulent flows. The existence of regions characterized by enhanced turbulence and dissipation, as observed, for instance, in the ocean, requires appropriate tools to assess how energy is transferred across the scales and at the same time locally in the physical space. After refining a classical coarse-graining procedure, here we investigate the feedback of extreme vertical velocity drafts on energy transfer and exchanges in subdomains of simulations of stably stratified flows of geophysical interest. Our analysis shows that vertical drafts are indeed able to trigger upscale and downscale energy transfers, strengthening the coupling between kinetic and potential energies at certain scales, depending on the intensity of the local vertical velocity.
\end{abstract}

\maketitle

\section{Introduction}
Stratified turbulence is widely investigated in the context of weather and climate studies. Indeed, the atmosphere and oceans are rotating and stratified flows, with their dynamics strongly influenced by the propagation of inertia-gravity waves from synoptic ($\sim 10^3$ km) to mesoscale ($\sim 10^2$ km). At the sub-mesoscale ($\sim 10$ km) motions are less constrained by the geostrophic balance \cite{mcwilliams}, and the interplay between turbulent fluctuations and propagating internal waves makes geophysical flows significantly different from homogeneous isotropic turbulence (HIT), even considering their dry dynamics only~\cite{laval_03,Alexakis2024}. Rotation and stratification affect the way energy is transferred in Fourier space~\cite{marino_14}, allowing for the onset of inverse~\cite{lindborg_06,marino_13} and bi-directional energy cascade in geophysical fluids~\cite{pouquet_13,marino_15,balwada_2022,Alexakis2024}. In the presence of stratification, the potential temperature is coupled with the velocity field, opening a channel for the exchange between kinetic and potential energy modes~\cite{billant_00}. A measure of the strength of gravity waves in stratified turbulent flows is provided by the Froude number, $Fr=\uptau_{\rm W_g}/\uptau_{\rm NL}$, defined as the ratio between the characteristic time associated to buoyancy, $\uptau_{\rm W_{g}}=1/N$ (where N is  Brunt-V\"{a}is\"{a}l\"{a} frequency, defined later), and the nonlinear time $\uptau_{\rm NL}=L_{\rm int}/U_{\rm rms}$, where $L_{\rm int}$ and $U_{\rm rms}$ represent the integral scale and the characteristic root mean square (RMS) velocity of the system, respectively. Due to the large number of scales involved, geophysical flows are often investigated with the help of models that parametrize the small-scale dynamics, such as in large-eddy simulations (LES)~\cite{Khani2015,Khani2020}; at the same time, the rapid growth of computational capabilities allowed to perform direct numerical simulations (DNS) at very large Reynolds numbers and, in general, using parameters compatible with that of the atmosphere and oceans~\cite{Rosenberg2015,deBruynKops2015,Petropoulos2024,Alexakis2024}.
The use of high-resolution DNS, allowed to show how both vertical velocity and potential temperature in stratified turbulent flows can develop large-scale intermittent events (thus at scales comparable with that of the mean flow)~\cite{rorai_14,feraco_18}, characterized by non-Gaussian statistics, as observed in the atmospheric boundary layer~\cite{mahrt_89,lyu_18}, stratosphere~\cite{imazio_22}, mesosphere~\cite{chau_21}, and oceans~\cite{dasaro_07}. 
The concept of intermittency in turbulence is broad, with intermittent phenomena being observed in a variety of frameworks in nature, on Earth as well as in the outer space~\cite{Burlaga,Castaing}. Generally associated with the departure of the small-scale field fluctuations statistics from gaussianity, it can indeed occur as well at large scales~\cite{k62}. 
Extensive parametric explorations presented in~\cite{feraco_18,Feraco2021} using DNS of the Boussinesq equations, demonstrated that powerful large-scale vertical drafts and temperature bursts occur in a certain range of Froude numbers of geophysical interest. These authors also proposed simplified models explaining how extreme events arise from resonant interactions between internal gravity waves and turbulent motions~\cite{rorai_14,Sujovolsky2019,Sujovolsky2020,marino_22}. 
The causal link between the emergence of extreme vertical velocity drafts, the fourth-order moment of the vertical velocity (namely, the kurtosis $K_w$), and the enhancement of local turbulence, dissipation and mixing in stratified flows was also established~\cite{feraco_18,pouquet_18,pouquet_19,marino_22}. In particular, vertical drafts proved to be essential for stratified turbulent flows to dissipate energy as efficiently as HIT flows~\cite{marino_22}, providing an explanation for the relation observed in the ocean between the intermittent emergence of localized turbulence pattern and the dissipation being concentrated in a relatively small portion of the global ocean volume~\cite{Fox-Kemper,Fontanet}. 
While the contribution of extreme vertical velocity drafts to the energetics of stratified turbulent flows has been so far assessed in terms of their feedback on the domain volume statistics \cite{marino_22}, their irregular emergence in space and time made it difficult to investigate how they may affect the spectral energy distribution at the location where they are detected. 
Classical three-dimensional Fourier transforms are in fact global operations, implying overall volume computations, thus averaging over regions whose dynamics can vary significantly due to the presence of these large-scale intermittent events. 
In order to investigate kinetic and potential energy transfers across scales in regions characterized by large values of the vertical velocity, we {used the well-established approach known as} coarse-graining ~\cite{eyink_09,Aluie2010,camporeale_18,Aluie_2018,Buzzicotti_2021} to three-dimensional DNS of the Boussinesq equations~\cite{Aluie2011,Zhou2022,Zhou2024}. {In particular, we refined} the implementation of {classical} spatial filters to obtain accurate estimates of the axisymmetric fluxes and investigate the possibility for extreme vertical velocity drafts to act as a local kinetic energy injection mechanism and/or to enhance exchanges between kinetic and potential energy.\\
The paper is organized as follows. Sec.~\ref{sec:drafts_lit}  briefly introduces the fluid framework under study and some of its features. Sec.~\ref{sec:Bssnsq_eqs} describes the space-filtered energy equations in the Boussinesq framework. Sec.~\ref{sec:DNS} provides an overview of the numerical framework used to perform the simulations analyzed. In Sec.~\ref{sec:role_of_drafts} we assess the feedback of the extreme vertical velocity drafts on local energy transfer and exchanges, as it occurs in localized regions of the physical space. In Sec.~\ref{sec:conclusions}, main results are summarized and further discussed.

\section{Large-scale intermittency in stratified geophysical flows}\label{sec:drafts_lit}

Turbulent {fluids} develop strong field gradients, a phenomenon known as internal (or small-scale) intermittency associated with {the emergence of} patches of dissipation distributed more or less homogeneously {in the domain flow}. {This classical form of intermittency is  responsible for the departure from Gaussianity of the statistics of the field fluctuations when small spatial or temporal increments are considered, affecting high-order statistical moments \cite{Leveque_1994}.} However, intermittency is not only present at the smallest scales. {As already mentioned,} in stratified geophysical flows strong fluctuations of the fields are in fact observed at {the scale} of the mean flow~\cite{mahrt_89,dasaro_07,lyu_18,Fontanet}. The large-scale intermittent behavior of the vertical velocity and temperature has recently been investigated in DNS of the Boussinesq equations, exploring a parameter space relevant for the atmosphere and the oceans, and allowing for the characterization of this phenomenon in terms of the interplay between internal gravity waves and turbulent motions. In particular, very large fluctuations in the vertical component of the velocity and potential temperature, diagnosed through the kurtosis of the fields, were observed at Froude numbers of order $10^{-2}$~\cite{feraco_18,Feraco2021}. By examining the kurtosis of the vertical velocity ($K_w$), a transition was found across values of $Fr$ of this order, as stratification strengthens, leading to heavy non-Gaussian tails of the probability distribution functions (PDFs). The existence of a resonant regime characterized by enhanced large-scale intermittency was invoked, based on a one-dimensional model proposed in~\citet{rorai_14} and \citet{feraco_18}, to explain the emergence of strong velocity and potential temperature field fluctuations, associated with localized overturning~\cite{pouquet_23},
enhanced mixing and dissipation~\cite{pouquet_18,pouquet_19,marino_22}. 

\section{Coarse-graining approach for stably stratified flows:the reduced energy fluxes}\label{sec:Bssnsq_eqs}
% \section{Coarse-graining approach for stably stratified flows: reduced energy fluxes}\label{sec:Bssnsq_eqs}

The rationale behind using the {coarse-graining} approach to investigate the feedback of large-scale vertical drafts on the flow fields is that the ``sub-{filter}'' scale energy transfer and exchange terms obtained from the filtered equations provide estimates analogous to the classical Fourier fluxes {with the advantage that} {they are} defined locally in the physical space~\cite{eyink_09,Aluie2009}, as shown in the following. This approach will therefore be used to investigate the point-wise energy transfer and exchange of (or between) kinetic and potential energy across the scales in stratified turbulent flows. A coarse-graining {procedure} was first employed in the context of large-eddy simulations (LES)~\cite{Leonard1975,germano1992,Meneveau2000} and more recently it has been successfully applied in a variety of studies to investigate the energy transfer in simulations of fluids~\cite{hellinger_21} and plasmas~\cite{Aluie2010,yang_17,CerriCamporealePOP2020,Manzini2022a,Manzini2022b} and also analyzing experimental data~\cite{Buzzicotti_2021,deleo_2022,Khatri2024}.
This work focuses on characterizing the contribution of the vertical velocity drafts to the transfer of kinetic and potential energy. Specifically, three-dimensional DNS of the Boussinesq equations, as reported below, will be analyzed:

\begin{align}
    \partial_t\,\uv\, 
    +\, \bomega\btimes\uv\, 
    =\,
    \,- & N\theta\ev_z\, 
    -\, \bnabla\Ptot\,
    +\, \nu\nabla^2\uv
    +\, \bb{F}_{\rm ext}\,,\label{eq:Bssnsq_flow}\\
    \partial_t\,\theta\, 
    +\, \big(\uv\cdot\bnabla\big)\theta\, 
    =\, 
    \,\, & Nw\, 
    +\, \kappa\nabla^2\theta\,,\label{eq:Bssnsq_theta}
\end{align}
where $\bb{u}$ is the velocity field (with $\bnabla\cdot\bb{u}=0$), $\theta$ the temperature fluctuations around a mean temperature $\theta_0$, $\bomega\doteq\bnabla\btimes\uv$  the flow vorticity, $N\doteq\sqrt{-(g/\theta_0)\partial\bar{\theta}/\partial z}$ the Brunt-V\"{a}is\"{a}l\"{a} frequency {($\bar{\theta}$ being the background linear temperature profile)}, $\gv\doteq-g\ev_z$ the gravity, $\mathcal{P}$ the total pressure\footnote{This scalar quantity includes the kinetic energy density (per unit mass), $|\bb{u}|^2/2$, as a consequence of rearranging the nonlinear term in the Navier-Stokes equation, $\big(\bb{u}\cdot\bnabla\big)\bb{u}=\bb{\omega}\btimes\bb{u}+\bnabla\big(|\bb{u}|^2/2\big)$.}, $w\doteq\bb{u}\cdot\bb{e}_z$ the component of the flow along the gravity direction, and $\bb{F}_{\rm ext}$ an external forcing applied to the velocity field. The parameters $\nu$ and $\kappa$ are kinematic viscosity and diffusivity, respectively. From the Boussinesq equations~\eqref{eq:Bssnsq_flow}--~\eqref{eq:Bssnsq_theta} it is straightforward to obtain evolution equations for kinetic and potential energies, $\Eu\doteq|\uv|^2/2$ and $\Etheta\doteq\theta^2/2$:

\begin{align}
    \partial_t\,\Eu\, 
    +\, \bnabla\cdot\big(\Ptot\uv-\nu\nabla\Eu\big)\, 
    =\, &
    \,- N\theta w\, +\, D_\nu\, +\, \epsilon_{\rm ext}\,,\label{eq:Bssnsq_flow_energy}\\
%    
    % & \nonumber\\
%    
    \partial_t\,\Etheta\, 
    +\, \bnabla\cdot\big(\Etheta\uv-\kappa\nabla\Etheta\big)\, 
    =\, & 
    \,\,N\theta w\, +\, D_\kappa\label{eq:Bssnsq_potential_energy}\,,
\end{align}
%%%%%%%%%%%%%%%%%%%%%%%%%%%%%%%%%%%%%%%%%%%%%
here the dissipation terms are {$D_\nu\doteq-\nu||\bSigma||^2$},  $||\bSigma||^2=\bSigma:\bSigma=\Sigma_{ij}\Sigma_{ji}$ being the square modulus of the strain tensor $\Sigma_{ij}\doteq\partial_iu_j$, and {$D_\kappa\doteq-\kappa|\bnabla\theta|^2$}. The external kinetic energy injection rate is $\epsilon_{\rm ext}\doteq\bb{F}_{\rm ext}\cdot\bb{u}$. The term $N\theta w$ that appears with opposite sign in both equations~\eqref{eq:Bssnsq_flow_energy} and~\eqref{eq:Bssnsq_potential_energy}, often referred to as buoyancy flux $B_f$~\cite{feraco_18}, represents a ``conversion'' term between the kinetic and potential energies. These two energies are coupled through the nonlinear interaction between vertical velocity and temperature fluctuations, $w$ and $\theta$, respectively.
By performing a spatial average over the whole fluid domain (operation denoted by $\langle\dots\rangle$), assuming vanishing fluxes at the boundaries, i.e., that $\langle\bnabla\cdot(\dots)\rangle=0$, the above energy equations read as

\begin{align}
    \partial_t\langle\Eu\rangle\,
    =\, &
    \,-\langle N\theta w\rangle\, +\, \langle D_\nu\rangle\, +\,\langle\epsilon_{\rm ext}\rangle\,,\label{eq:Bssnsq_flow_energy_averaged}\\
%    
    % & \nonumber\\
%    
    \partial_t\,\langle\Etheta\rangle\, 
    =\, & 
    \,\,\langle N\theta w\rangle\, +\, \langle D_\kappa\rangle\label{eq:Bssnsq_potential_energy_averaged}\,.
\end{align}
%%%%%%%%%%%%%%%%%%%%%%%%%%%%%%%%%%%%%%%%%%%%%
The term $\langle N\theta w\rangle$ is responsible for the exchanges between the two types of energy, $\Eu$ and $\Etheta$, and it disappears when summing up equations~\eqref{eq:Bssnsq_flow_energy_averaged} and~\eqref{eq:Bssnsq_potential_energy_averaged} to obtain an equation for the total energy $\mathcal{E}_{\rm tot}=\Eu+\Etheta$ (conserved in case of vanishing dissipation, $D_\nu=D_\kappa=0$, and no external energy injection, $\langle\epsilon_{\rm ext}\rangle=0$). 

%============================
\begin{figure}[t]
\centering
\includegraphics[width=0.6\columnwidth]{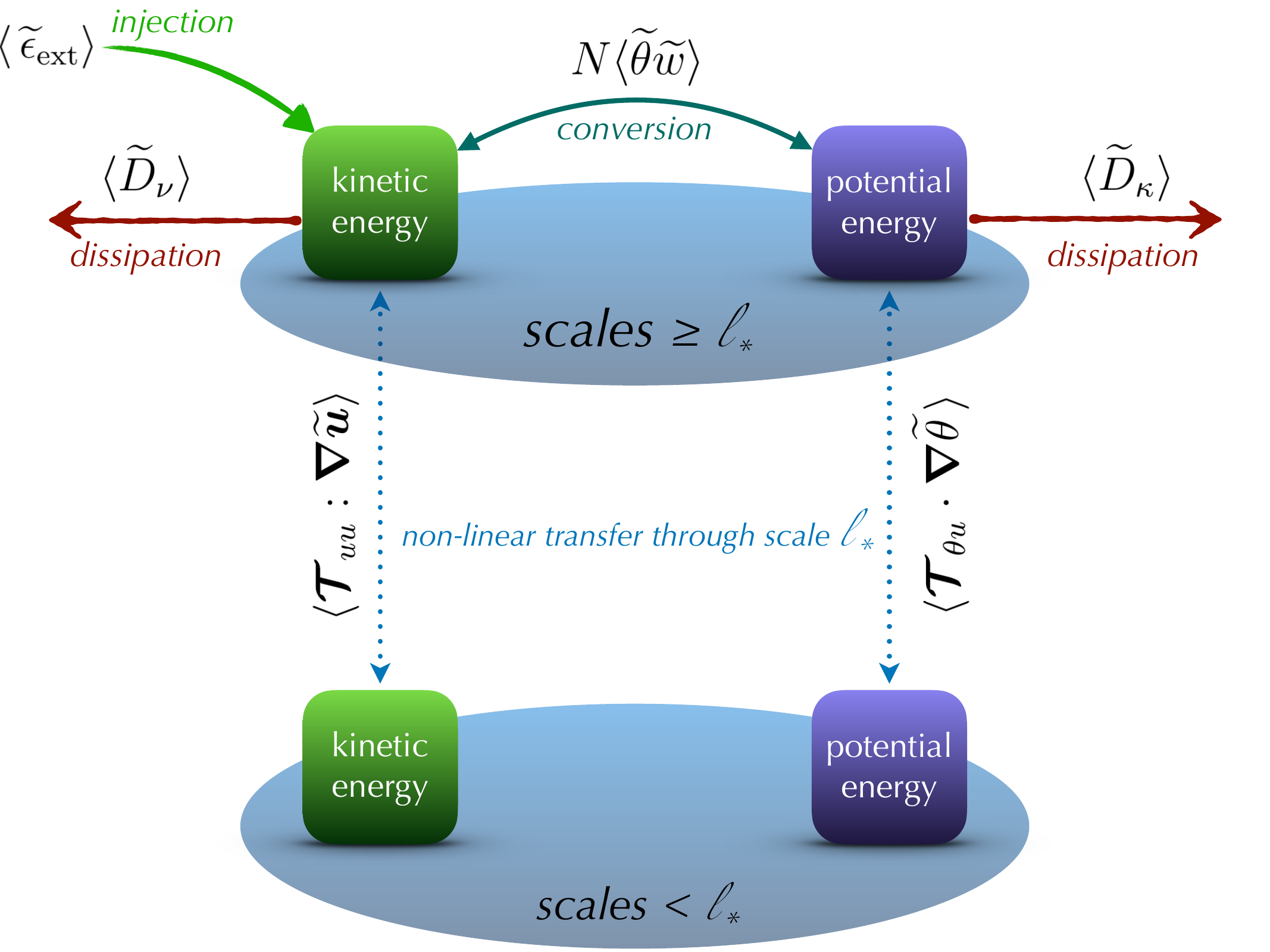}
\caption{Schematics of the channels resulting from the space-averaged energy equations for the filtered flux terms in Eqs.~\eqref{eq:Bssnsq_flow_energy_filtered_averaged} and~\eqref{eq:Bssnsq_potential_energy_filtered_averaged}. $\ell_*\sim 1/k^*$ denotes the characteristic scale of the applied low-pass filter.}
\label{fig:cartoon}
\end{figure}
%============================
\subsection{Filtered energy equations}\label{subsec:Bssnsq_energy_eqs}
Following the approach detailed in~\citet{CerriCamporealePOP2020} and references therein, we apply the {coarse-graining} technique to equations (\ref{eq:Bssnsq_flow})--(\ref{eq:Bssnsq_theta}), deriving the evolution equations for the ``large-scale filtered'' kinetic and potential energies.
This procedure consists in applying a low-pass filter at the cutoff scale $\ell_*$ and then restoring a filtered version of the energy equations, analogously to those in~\eqref{eq:Bssnsq_flow_energy} and~\eqref{eq:Bssnsq_potential_energy}, describing the evolution of the large-scale (i.e., $\ell \ge \ell_*$) kinetic and potential energies.
The filtered terms stemming from the nonlinear terms in the Boussinesq equations will be called ``sub-{filter scale}'' terms (or sub-scale terms), which explicitly represent the energy transfer between (all) the scales $\ell\geq\ell_*$ and (all) the scales below the filter $\ell<\ell_*$. 
% \rf{\sout{The direction of this transfer is encoded in the sign of these terms: the energy being transferred to scales smaller than $\ell_*$ is seen as a sink, thus a negative sign is associated with this transfer, and vice versa. Also,}}
This procedure does not assume the locality of the interactions in the Fourier space, so that the ``sub-{filter scale}'' terms account for (multiple) couplings between any of the scales smaller than $\ell_*$ with any of the scales larger than $\ell_*$. 
The space-filtered version of a vector field $\bb{v}(\bb{x},t)$ will be denoted as $\widetilde{\bb{v}}(\bb{x},t)$, and is defined as the convolution of $\bb{v}$ with a filter function $G$:
%%%%%%%%%%%%%%%%%%%%%%%%%%%%%%%%%%%%%%%
\begin{equation}\label{eq:filter-def-1}
 \widetilde{\bb{v}}(\bb{x},t)\,\doteq\,\int_{{\cal V}}G(\bb{x}-\bb{\xi})\bb{v}(\bb{\xi},t)\rmd^3\bb{\xi}\,,
\end{equation}
%%%%%%%%%%%%%%%%%%%%%%%%%%%%%%%%%%%%%%%
where ${\cal V}$ is the entire spatial domain. 
The filtering operation in~\eqref{eq:filter-def-1} is such that it commutes with differentiation in time and space:
%%%%%%%%%%%%%%%%%%%%%%%%%%%%%%%%%%%%%%%
\begin{equation}\label{eq:filter-def-2}
 \widetilde{\,\partial_t\,\bb{v}\,}\,=\,\partial_t\,\widetilde{\bb{v}}\,\qquad\mathrm{and}\qquad\,
 \widetilde{\grad\cdot\bb{v}}\,=\, \grad\cdot\widetilde{\bb{v}}\,.
\end{equation}
%%%%%%%%%%%%%%%%%%%%%%%%%%%%%%%%%%%%%%%
However, {according to the Germano identity~\cite{germano1992}}, i.e., $\widetilde{\,\bb{vv}\,}\neq\widetilde{\bb{v}}\,\widetilde{\bb{v}}$, {we can define} the corresponding ``sub-{filter scale''} term as
%%%%%%%%%%%%%%%%%%%%%%%%%%%%%%%%%%%%%%%
\begin{equation}\label{eq:subgrid-def}
 \bTvv\,\doteq\,\widetilde{\,\bb{vv}\,}\,-\,\widetilde{\bb{v}}\,\widetilde{\bb{v}}\,.
\end{equation}
%%%%%%%%%%%%%%%%%%%%%%%%%%%%%%%%%%%%%%%
If $\ell_*$ is the cutoff scale associated with the filter (as discussed above), then $\bTvv$ describes the coupling of all the scales $\ell\geq\ell_*$ to all the scales $\ell<\ell_*$ due to the nonlinear term $\bb{vv}$.
With the above definitions in mind, one obtains the filtered Boussinesq equations by applying the filtering procedure to equations~\eqref{eq:Bssnsq_flow}--\eqref{eq:Bssnsq_theta} and appropriately rewriting the nonlinear terms:

\begin{align}
    \partial_t\,\wtuv\, 
    +\, \wtbomega\btimes\wtuv\, 
    +\, \bTomegacrossu\,
    =\,- & N\,\wttheta\ev_z\, 
    -\, \bnabla\wtPtot\,
    +\, \nu\nabla^2\wtuv\, +\,\widetilde{\bb{F}}_{\rm ext}\,,\label{eq:Bssnsq_flow_filtered}\\
%
    % & \nonumber\\
%
    \partial_t\,\wttheta\, 
    +\, \bnabla\big(\wttheta\,\wtuv+\bTthetau\big)\, 
    =\,\,\, & N\wtw\,
    +\, \kappa\nabla^2\wttheta\,,\label{eq:Bssnsq_theta_filtered}
\end{align}
where $\bTomegacrossu\doteq\reallywidetildeB{\bomega\btimes\uv}-\wtbomega\btimes\wtuv\,$ and $\bTthetau\doteq\widetilde{\theta\uv}-\wttheta\,\wtuv\,$. From equations~\eqref{eq:Bssnsq_flow_filtered} and~\eqref{eq:Bssnsq_theta_filtered} one can derive the expression for the filtered kinetic and potential energy by taking the scalar product of~\eqref{eq:Bssnsq_flow_filtered} with $\wtuv$ and multiplying~\eqref{eq:Bssnsq_theta_filtered} by $\wttheta$, that reads as
\footnote{Since the total pressure $\mathcal{P}$ contains the contribution from the kinetic energy density, $|\uv|^2/2$, the term $\wtPtot-\mathrm{tr}[\bTuu]/2$ in~\eqref{eq:Bssnsq_flow_energy_filtered} corresponds to $\widetilde{p}+|\wtuv|^2/2=\widetilde{p}+\wtEu$. This is a consequence of the {definition} $\widetilde{\,|\uv|^2}=|\wtuv|^2+\mathrm{tr}[\bTuu]$.}

\begin{flalign}
    \partial_t\,\wtEu\,
    +&\, \bnabla\cdot\left[\left(\wtPtot-\frac{\mathrm{tr}[\bTuu]}{2}\right)\wtuv\,+\,\bTuu\cdot\wtuv-\nu\nabla\wtEu\right]\,
    =\,\,-N\wttheta\wtw\,
    +\,\bTuu:\bnabla\wtuv\,
    +\,\wtD_\nu\,
    +\, \widetilde{\epsilon}_{\rm ext}\,,\label{eq:Bssnsq_flow_energy_filtered}\\
%
    % & \nonumber\\
%
    \partial_t\,\wtEtheta\,
    +&\, \bnabla\cdot\left(\wtEtheta\wtuv\,+\,\bTthetau\wttheta\,-\kappa\nabla\wtEtheta\right)\,
    =\,\,\,N\wttheta\wtw\,
    +\,\bTthetau\cdot\bnabla\wttheta\,
    +\,\wtD_\kappa\,,\label{eq:Bssnsq_potential_energy_filtered}
\end{flalign}
%%%%%%%%%%%%%%%%%%%%%%%%%%%%%%%%%%%%%%%%%%%%%
where {$\wtD_\nu\doteq-\nu||\wtbSigma||^2$}, {$\wtD_\kappa\doteq-\kappa|\bnabla\wttheta|^2$}, and $\widetilde{\epsilon}_{\rm ext}\doteq\widetilde{\bb{F}}_{\rm ext}\cdot\wtuv$ are the filtered dissipation terms and kinetic energy injection rate in~\eqref{eq:Bssnsq_flow_energy}--\eqref{eq:Bssnsq_potential_energy}, respectively. In rewriting the term $\wtuv\cdot\bTomegacrossu$ we used the incompressibility condition, $\bnabla\cdot\wtu=0$, along with the fact that the sub-{filter} term arising from $\bomega\btimes\uv$ can be rewritten as $\bTomegacrossu=\bnabla\cdot[\bTuu-(\mathrm{tr}[\bTuu]/2)\bb{I}]$, $\bTuu\doteq\widetilde{\,\bb{uu}\,}-\wtuv\wtuv$ being the sub-{filter scale} {turbulent} stress tensor of the flow (while $\mathrm{tr}[\bTuu]$ is its trace), and $\bb{I}$ the identity tensor. We remind the reader that the symbol ``:'' is the tensor scalar product, i.e., $\bTuu:\bnabla\wtuv=(\mathcal{T}_{uu})_{ij}\,\partial_j\widetilde{u}_i$.
Analogously to equations~\eqref{eq:Bssnsq_flow_energy_averaged}--\eqref{eq:Bssnsq_potential_energy_averaged}, performing volume averages of~\eqref{eq:Bssnsq_flow_energy_filtered}--\eqref{eq:Bssnsq_potential_energy_filtered} leads to

\begin{align}
    \partial_t\,\langle\wtEu\rangle\,
    =\, &
    \,- N\langle\wttheta\wtw\rangle\,
    -\,\langle\mathcal{S}_u\rangle\,
    +\,\langle\wtD_\nu\rangle\,
    +\,\langle\,\widetilde{\epsilon}_{\rm ext}\rangle\,,\label{eq:Bssnsq_flow_energy_filtered_averaged}\\
%
    % & \nonumber\\
%
    \partial_t\,\langle\wtEtheta\rangle\,
    =\, & 
    \,\,N\langle\wttheta\wtw\rangle\,
    -\,\langle\mathcal{S}_\theta\,\rangle\,
    +\,\langle\wtD_\kappa\rangle\,,\label{eq:Bssnsq_potential_energy_filtered_averaged}
\end{align}
%%%%%%%%%%%%%%%%%%%%%%%%%%%%%%%%%%%%%%%%%%%%%
where we defined the sub-{filter} terms $\mathcal{S}_u\doteq-\bTuu:\bnabla\wtuv$ and $\mathcal{S}_\theta\doteq-\bTthetau\cdot\bnabla\wttheta$ for brevity.
From equation~\eqref{eq:Bssnsq_flow_energy_filtered_averaged} one infers that the transfer rate of kinetic energy through a scale $\ell_*$ stems from the interaction between the strain tensor at scales $\ell\geq\ell_*$, $\wtbSigma=\bnabla\wtuv$, and the sub-{scale} {turbulent} stress, $\bTuu$.
Similarly, equation~\eqref{eq:Bssnsq_potential_energy_filtered_averaged} shows that the transfer rate of potential energy through $\ell_*$ depends on the interaction of the sub-{scale} heat flux, $\bTthetau$, with the gradient of temperature fluctuations at scales $\ell\geq\ell_*$, $\bnabla\wttheta$.
Potential and kinetic energy channels are coupled by $N\neq0$, which allows the conversion rate between the two energy forms through the nonlinear term involving temperature and vertical fluctuations at scales $\ell\geq\ell_*$, i.e., $N\langle\wttheta\wtw\rangle$. A schematic view of the global (i.e., space-averaged) dynamics of the kinetic and potential energy channels described by equations~\eqref{eq:Bssnsq_flow_energy_filtered_averaged}--\eqref{eq:Bssnsq_potential_energy_filtered_averaged} is depicted in Figure~\ref{fig:cartoon}. {Here, we omitted the energy conversion term at scales smaller than the cut-off scale (e.g., see~\cite{Aluie2017,Zhao2022}) as we focus on the large-scale energetics.} {It is also worth noticing that a non-uniform stratification can strongly modify the energy transferred across a single channel via the terms $\TN[\theta]$, and $\TN[w]$, as well as the exchanges between $\wtEu$ and $\wtEtheta$ (since $\langle\wtN\wttheta\wtw\rangle\neq N\langle\wttheta\wtw\rangle$).}\\
Summing up the equations for filtered kinetic and potential energies, one obtains the scale-by-scale conservation equation for the filtered total energy, $\widetilde{\mathcal{E}}_{\rm tot}\doteq\wtEu+\wtEtheta$, in which the conversion terms $N\wttheta\wtw$ cancel out and the transfer rate of total energy across the scale $\ell_*$ is given by $\mathcal{S}_{\rm tot}=\mathcal{S}_u+\mathcal{S}_\theta$. Finally, it is worth reminding that at any fixed $\ell_*$, if a sub-{scale} term $\mathcal{S}$ is positive (negative), then $\mathcal{S}$ represents a sink (source) term as seen by the energy reservoir $\widetilde{\mathcal{E}}$ at scales $\ell\geq\ell_*$, and thus the energy is being transferred to (from) scales $\ell<\ell_*$ from (to) scales $\ell\geq\ell_*$. This sign convention for the sub-{scale} terms is consistent with the classical Fourier energy flux (see Section~\ref{subsec:Fourier_vs_spacefilter}).
\begin{table}[t]
\centering
\resizebox{0.4\columnwidth}{!}{%
\begin{tabular}{cc|cccccccccccccccccc}
\hline \hline
Run & & & $k_F$ & & $N$ & & $Re$ & & $Fr$  & & $R_B$ & & $n_p$\\ 
% & & $k_\eta$ & & $k_b$ & & $k_{Oz}$\\ 
\hline
I   & & & 20    & & 8   & &  97    & &   0.128    & &   1.59   & & $512^3$ 
% & & & & & 
\\ \hline
% II  & & & [8,9] & & 8   & &  455    & &  0.104     & &  4.92      \\ \hline
II & & & $2.5$ & & 8   & &  3800    & & 0.076 & &  22.1    & & $512^3$ 
% & & & 7 & & 66
\\ \hline \hline
\end{tabular}
}
\caption{Relevant parameters of the DNS analyzed: $k_F$ is the forcing wave number, 
$N$ the Brunt-V\"{a}is\"{a}l\"{a} frequency, $Re$ the Reynolds number, $Fr$ the Froude number and $R_B$ the buoyancy Reynolds.}
\label{tab:runs}
\end{table}
In order to include the conversion between the two energy channels as possible source/sink terms for each other energy reservoir, we define the ``conservative outflux'' of kinetic (potential) energy from scales $\ell\geq\ell_*$ as $\Phi_u\doteq\mathcal{S}_u+N\wttheta\wtw$ ($\Phi_\theta\doteq\mathcal{S}_\theta-N\wttheta\wtw$); this allows a direct comparison between the volume-average sub-{scale} terms computed here and the scale-to-scale Fourier energy flux. Note that the conversion terms describe energy conversions occurring entirely at scales $\ell\geq\ell_*$, whereas only the sub-{scale} terms properly describe the energy transfer through scales, i.e., the transfer between the ``scale domains'' $\ell\geq\ell_*$ and $\ell<\ell_*$ that is passing through the cutoff $\ell_*$ (the sign of the sub-{scale} term giving the direction of this transfer). Given that the conversion terms cancel out for the total energy, it readily follows that $\Phi_{\rm tot}=\mathcal{S}_{\rm tot}$. 
In general, the final form of the filtered Boussinesq equations~\eqref{eq:Bssnsq_flow_filtered}--\eqref{eq:Bssnsq_theta_filtered},  hence of the corresponding filtered energy equations~\eqref{eq:Bssnsq_flow_energy_filtered_averaged}--\eqref{eq:Bssnsq_potential_energy_filtered_averaged}, are independent of the particular choice made for the filtering kernel $G$ (e.g. low-pass, top-hat, Gaussian, or other filter shapes).
Here we perform convolutions between the physical variables (i.e., velocity and temperature fluctuations) and the Butterworth filter, defined in the Fourier space as $G^{(n)}(k)=1/[1+(k/k^*)^{2n}]$, with $n=4$ and $k^*$ the characteristic wave number above which fluctuations are filtered out, thus corresponding to a low-pass filter. {Focusing on the large-scale energy transfer equations, this choice seems appropriate} {for our analysis, though it is worth mentioning that such filter may lead to point-wise negative values of the small-scales kinetic energy, defined as $\mathrm{tr}\left[\bTuu\right]=\widetilde{|\uv|^2}-|\widetilde{\uv}|^2$.}\\
% \NOTE{[Raffaello, controlla per favore se abbiamo valori negativi]}
The choice of an isotropic filter acting on circular or spherical shells (defined by the wave number modulus only, $k$) is straightforward for the analysis of homogeneous and isotropic flows, either in two or three dimensions~\cite{camporeale_18}. However, stratified turbulent flows are anisotropic and a reasonable option is to implement a spatial filters analogous to classical Fourier integrations through planes or cylindrical shells, as when the computation of parallel and perpendicular energy fluxes is operated in the Fourier space (see~\ref{subsec:Fourier_vs_spacefilter} for the definition). 
It will be shown in the next section how the correspondence between reduced fluxes in the Fourier space and sub-{scale} flux terms is achieved by modifying the symmetry properties of the filtering kernel $G^{(4)}(k_{\perp,\parallel})=1/[1+(k_{\perp,\parallel}/k^*_{\perp,\parallel})^{8}]$. In particular, we obtain parallel and perpendicular integrated sub-{scale} terms, $\mathcal{S}(k_\parallel)$ and $\mathcal{S}(k_\perp)$, respectively , assuming the filters $G^{(4)}(k_\parallel)$, with $k_\parallel=|k_z|$ (where gravity is along the parallel direction), and $G^{(4)}(k_\perp)$, with $k_\perp=\left(k_x^2+k_y^2\right)^{1/2}$.\\ The choice to perform filters along parallel and perpendicular directions (with respect to gravity) in the physical space~\cite{Aluie2023}, is  motivated as well by the fact that numerous previous studies highlighted a different behavior of the energy transfer when fluxes result from integrations along different directions in the Fourier space, in stratified~\cite{Billant2001} and rotating stratified turbulent flows~\cite{marino_14}. On the other hand, attention must be paid in analyzing reduced energy fluxes in turbulent flows; in particular, using DNS of the Boussinesq equations reproducing the planetary atmospheres in a realistic parameter space, it has been recently shown that partial fluxes may not capture the actual energy cascade rate in geophysical flows~\cite{Alexakis2024}. 
\begin{figure}[t]
\centering
\includegraphics[width=\textwidth]{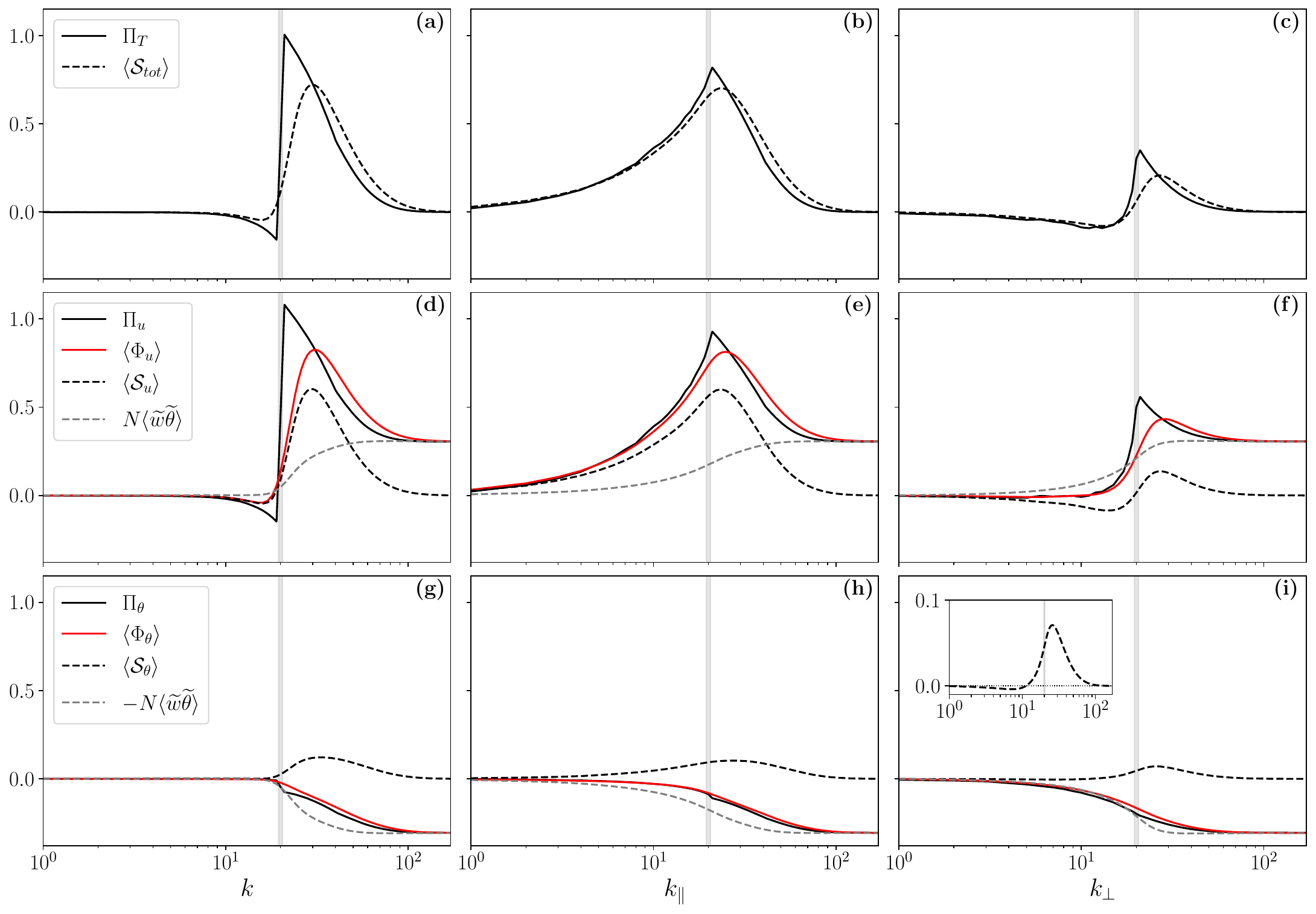}
\caption{Panels (a)--(c): comparison between isotropic (left), parallel (center) and perpendicular (right) scale-to-scale Fourier total energy flux $\Pi_{tot}$ (black solid lines) and the total energy sub-{scale} flux terms (black dashed lines) for run II. In panels (d)--(f) and (g)--(i) the same comparison is proposed for the kinetic and potential energy fluxes, respectively, and the individual flux terms. For the single energy channels, $\widetilde{\Phi}$ (red lines) is the sum of the cross-scale term $\langle\mathcal{S}_{u,\theta}\rangle$ and $\mp N\langle\wttheta\wtw\rangle$, representing the conservative terms on the right-hand side of Eqs.~\eqref{eq:Bssnsq_flow_energy_filtered_averaged}--~\eqref{eq:Bssnsq_potential_energy_filtered_averaged}. The channel-conversion terms alone $\mp N\langle\wttheta\wtw\rangle$ are also shown (gray dashed lines). The inset in panel (i) shows a detail of the potential energy sub-{scale} term. A vertical dashed area denoting $k_F=20$ is provided.}
\label{fig:run_S20}
\end{figure}

\section{Direct Numerical Simulations}\label{sec:DNS}
The Boussinesq equations~\eqref{eq:Bssnsq_flow}--\eqref{eq:Bssnsq_theta} are solved in a triply periodic cubic box of size $L_0=2\pi$, discretized on a uniform grid, using the highly parallelized pseudo-spectral code GHOST~\cite{GHOST11,GHOST2020}.
An external random forcing acting on the velocity field only (the temperature field is not forced) continuously injects energy in an isotropic wave number shell $k_F=2\pi/L_F$.
The main governing parameters of the flow are the Reynolds number $Re=U_{\rm rms}L/\nu$, and the Froude number $Fr=U_{\rm rms}L_{\rm int}/N$. Here the integral scale $L_{\rm int}$ is taken as the scale at which the external forcing is applied, i.e., $L_{\rm int}=L_F$. Combining $Re$ and $Fr$ parameters one can define the so-called buoyancy Reynolds number $R_B=Fr^2Re$. 
Table~\ref{tab:runs} collects the relevant parameters used in {two} simulations {of stratified turbulent flows,} analyzed in this study, performed on grids of $512^3$ points, varying the forcing wave number. In particular, run I is forced at intermediate scales, $k_F=20$, while kinetic energy is injected at large scale in run II, $k_F\approx 2.5$.  
Run I, characterized by weaker turbulence was used to test the filters design, comparing the sub-scale transfer terms in the equations with the classical reduced fluxes computed in the Fourier space along the perpendicular ($k_\perp$) and parallel ($k_\parallel$) directions as in~\citet{marino_14}. Run II is {instead} the same as the one thoroughly analyzed in~\citet{marino_22}, where the feedback of the extreme vertical drafts on global spectral properties and dissipation was explored. 
In the present work the space filtering approach allows to extend the analysis in~\citet{marino_22} by exploring how vertical drafts affect locally in the physical space by-scale distributions and exchanges dynamics in stably stratified flows in {presence} of large-scale intermittency.

\subsection{Anisotropic fluxes vs sub-filter scale terms}\label{subsec:Fourier_vs_spacefilter}
To better interpret the information {stemming from implementations of the coarse-graining} approach, in this section we compare sub-{scale} terms with the fluxes computed with the usual Fourier analysis over the entire domain volume. {In particular, this is done for the case of a stratified flow forced at intermediate scale to benchmark the capability of our algorithm to capture energy transfers towards both large and small scales, in the different directions of the Fourier space}. The anisotropic {total} energy transfer can be obtained from the axisymmetric transfer function~\cite{marino_14,Alexakis2024}, 

\begin{equation}
% \begin{aligned}
    \tau_T(k_{\perp},k_{\parallel})\, 
    \doteq\, 
    \int \left[\widehat{\bb{u}}_{\bf k}\cdot\reallywidehat{\left(\bb{u}\cdot\bnabla\bb{u}\right)}_{\bf k}^* + \widehat{\theta}_{\bf k}\cdot\reallywidehat{\left(\bb{u}\cdot\bnabla\theta\right)}_{\bf k}^*\right]k_\perp\,{\rm d}\phi +\, {\rm c.c.}
% \end{aligned}
\end{equation}\label{eq:def_t_kperp-kpara}

which can be also defined in terms of spherical coordinates as $\tau(k,\Theta)$, indicating the isotropic flux; in Eq.~\eqref{eq:def_t_kperp-kpara}, the hat $\widehat{(\dots)}_{\bf k}$ denotes the Fourier coefficient at scale $\mathbf{k}$, both $^*$ and c.c. stand for complex conjugate, and $\phi$ is the azimuthal angle (defined with respect to the $x$ axis; the parallel direction that defines $k_\|$ is the $z$ axis instead). {Analogously, transfer functions can also be defined for the kinetic and potential energy separately,}

\begin{align}
% \begin{aligned}
    \tau_u(k_{\perp},k_{\parallel})\, 
    &\doteq\, 
    \int \widehat{\bb{u}}_{\bf k}\cdot\reallywidehat{\left(\bb{u}\cdot\bnabla\bb{u}\right)}_{\bf k}^* k_\perp\,{\rm d}\phi +\, {\rm c.c.}\,,\label{eq:transfer_f_kin}\\
    \tau_\theta(k_{\perp},k_{\parallel})\, 
    &\doteq\, 
    \int \widehat{\theta}_{\bf k}\cdot\reallywidehat{\left(\bb{u}\cdot\bnabla\theta\right)}_{\bf k}^* k_\perp\,{\rm d}\phi +\, {\rm c.c.}
    \label{eq:transfer_f_pot}
% \end{aligned}
\end{align}

By integrating Eq.~\eqref{eq:def_t_kperp-kpara} over spheres, planes, and cylinders in Fourier space, we obtain respectively,

\begin{align}
    T_\alpha(k)\,
    &=\,
    \int \tau_\alpha(k,\Theta)\, k{\rm d}\Theta\,,\label{eq:def_T_k}\\
    T_\alpha(k_{\parallel})\,
    &=\,
    \int \tau_\alpha(k_{\perp},k_{\parallel})\, {\rm d}k_{\perp}\,,\label{eq:def_T_kpara}\\
    T_\alpha(k_{\perp})\,
    &=\,
    \int \tau_\alpha(k_{\perp},k_{\parallel})\, {\rm d}k_{\parallel}\,,\label{eq:def_T_kperp}
\end{align}
with $\alpha \in [T,u,\theta]$.
The integration of these fluxes leads to  Eq.~\ref{eq:def_Pi_T} which  represents the isotropic, parallel, and perpendicular (with respect to the direction of gravity) scale-to-scale energy flux, with $k_i \in [k,k_\perp,k_\parallel]$, respectively.

\begin{align}
    \Pi_\alpha(k_i)\,
    &=\,
    -\int_0^{k_i} T_\alpha(k_i')\, {\rm d}k_i'\,,\label{eq:def_Pi_T}
\end{align}

Both Eq.~\eqref{eq:def_Pi_T} and the sub-scale terms provide the global-in-scale energy flux through $k^*$, mediated by all the possible couplings between wave numbers with $k<k^*$ and those with $k>k^*$ (i.e., an ``all-to-all'' transfer). Possible discrepancies between {these flux estimates} are due to the fact that the non-sharp convolution kernel (in the spectral space) {of the coarse-graining procedure} represents a smoother version of the sharp spectral filter adopted in the Fourier analysis~\cite{Rivera2014}.

\begin{figure}[t]
\centering
\includegraphics[width=0.45\columnwidth]{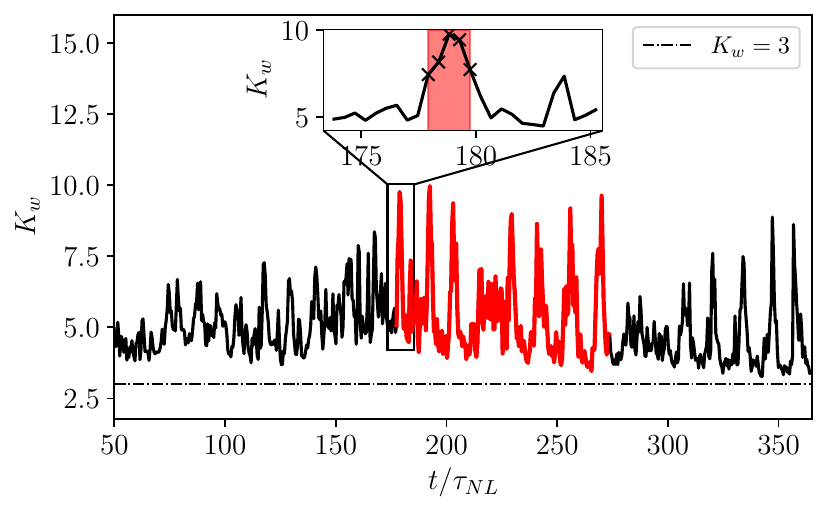}
\caption{Temporal evolution of kurtosis $K_w$ of the vertical velocity run II. The red portion of the curve corresponds to the interval analyzed in Sec.~\ref{subsec:time_anlysis}, the red shaded area is instead analyzed in Sec.~\ref{sec:role_of_drafts} and corresponds to $\sim 2\uptau_{NL}$; the same interval is evidenced in the inset. The horizontal black dash-dotted line is the Gaussian reference value for the kurtosis $K_w=3$.}
\label{fig:ch4_Kw_run_N8}
\end{figure}

{The comparison {between the outputs of the two procedures is proposed here for a stably stratified flow} Run II, {simulated on a grid of $512^3$ points, in which the} energy is injected at $k_F=20$ (Run I in Tab.~\ref{tab:runs}).}
The spectral transfer in a simulation with parameters similar to run I, with the same forcing mechanism, has been analyzed in~\citet{marino_14}, where the different components of the energy transfer (isotropic, perpendicular and parallel) have been characterized showing a different behavior in the presence of gravity and/or rotation. The results presented in~\citet{marino_14} show that in this particular setup the isotropic flux $\Pi_T(k)$ is almost zero for $k < k_F$, indicating that almost no energy {is transferred from the forcing scale} across spheres (in Fourier space) toward small{er} wave numbers in purely stratified turbulent flows, {or that no isotropic inverse cascade is detected in stratified turbulent flows in the absence of rotation}. On the other hand, the parallel flux $\Pi_T(k_{\parallel})$ {was observed to be} positive and dominant for all wave numbers, {indicative of a direct anisotropic transfer towards larger parallel wave numbers}. Completely different is the behavior of the perpendicular component of the flux $\Pi_T(k_{\perp})$, showing a range with negative values for $k < k_F$, indicating an inverse energy transfer ({thus towards larger scales}), and a positive flux for $k > k_F$ . Here, we {meant} to check whether this peculiar behavior {of purely stratified turbulent flows} can be captured by our implementation of the {coarse-graining} approach.\\
The comparison between the two techniques is reported in Fig.~\ref{fig:run_S20}. In panels (a)--(c) we show the total transfer computed with the Fourier method $\Pi_T(k_{i})$ (Eq.~\eqref{eq:def_Pi_T}, solid black line) and with the sub-{scale} terms $\langle\mathcal{S}_{\rm tot}\rangle=\langle\mathcal{S}_u+\mathcal{S}_{\theta}\rangle$ (black dashed line). Panels (d)--(f) and (g)--(i) show the energy flux associated with a single energy channel, i.e., kinetic (middle row) and potential (bottom row), respectively. For the latter cases, the Fourier flux (black line) is compared with the sum of the conservative terms on the right-hand side of Eqs.~\eqref{eq:Bssnsq_flow_energy_filtered_averaged}--\eqref{eq:Bssnsq_potential_energy_filtered_averaged}, i.e., $\widetilde{\Phi}_u=\mathcal{S}_u+N\wttheta\wtw$ and $\widetilde{\Phi}_\theta=\mathcal{S}_\theta-N\wttheta\wtw$, respectively (red lines). However, we also highlight the trend of the single terms composing the conservative flux i.e., $\mathcal{S}_{u,\theta}$ (black dashed line) and $N\wttheta\wtw$ (gray dashed lines).
From all the panels in Fig.~\ref{fig:run_S20} is evident the good agreement between the two approaches, both at large and small scales and for all the components. The discrepancy is more significant around the forcing wavenumber $k_F$ for the intrinsic difference between the {coarse-graining} approach and the Fourier analysis~\cite{Rivera2014}.
For $k>k_F$ the energy fluxes always indicate a downscale transfer of total energy, panels (a)--(c), with a modulation of the intensity going from the isotropic to the perpendicular component of the total energy flux. 
By looking at the perpendicular transfers, panel (c), the behavior previously described in terms of total Fourier energy flux is correctly recovered with the space filtering technique, showing an inverse transfer at scale $k_{\perp} < k_F$ and a direct transfer in the range $k_{\perp} > k_F$, with an inversion point with almost zero net flux at $k_{\perp} \sim k_F$.\\
Such a good agreement is also obtained for the single energy channels, in panels (f) and (i). In this case, some interesting features emerge from the analysis with the space filtering approach, in particular the role of the conversion term $N\langle\wttheta\wtw\rangle$ (dashed gray line in panels (d)--(i)), indicating the conversion from kinetic to potential energy if positive and vice-versa if negative, at scales $k < k_*$. Indeed, we can see from panel (f), for instance, how this term becomes the dominant contribution to the perpendicular flux at $k_{\perp} \gtrsim 70$, where $\langle\Phi_u\rangle\approx N\langle\wttheta\wtw\rangle$: this means that kinetic energy is almost totally converted into potential at small scales (let us remind that when $N\langle\wttheta\wtw\rangle$ is positive, it represents a sink term for $\langle\wtEu\rangle$ and a source term for $\langle\wtEtheta\rangle$ (cf. Eqs.~\eqref{eq:Bssnsq_flow_energy_filtered_averaged}--\eqref{eq:Bssnsq_potential_energy_filtered_averaged}), and this likely explains why we observe an upscale potential energy flux $\Pi_\theta$ at any $k>k_F$.
{This is also consistent with the fact that the total energy flux (computed as $\langle \mathcal{S}_{tot}\rangle$ or as $\Pi_T$), goes to zero at small scales, i.e there is no net {direct} {energy transfer} in this range, just a small-scale kinetic-to-potential energy conversion (plus dissipation---not shown here). By looking at the potential energy transfer {terms $\Pi_\theta$, $\langle \mathcal{S}_\theta\rangle$, and $\langle \Phi_\theta \rangle$ in} panel (i), one sees that the negative values {of the potential energy flux} are, in this case, not indicative of an inverse transfer of potential energy throughout the whole range of wave numbers{, and the scale-to-scale transfer is instead dominated by the kinetic-to-potential conversion term $N\langle\wttheta\wtw\rangle$}. In fact, the potential energy channel is not forced externally, this is only fed by the conversion of kinetic energy $\langle\Phi_{\theta}\rangle\approx N\langle\wttheta\wtw\rangle$ at each scale. The transfer of potential energy mediated by the nonlinear term $\langle \mathcal{S}_\theta\rangle$, in fact, still exhibits simultaneously positive (direct transfer) and negative (inverse transfer) values {(see inset in panel i of Fig.~\ref{fig:run_S20})}, although in this case the inversion scale -- i.e., the scale at which $\langle \mathcal{S}_\theta\rangle$ changes sign -- is not exactly at $k_F$, but at slightly larger scales (around $k_\perp\approx 10$, see inset).} The behavior of the potential energy transfer for the three components, panels (g)--(i), is pretty much the same, with an almost zero flux at $k<k_F$ and a negative transfer dominated by the conversion term at scale $k>k_F$.

\section{Local energy transfer and exchanges triggered by vertical drafts}\label{sec:role_of_drafts}
\begin{figure}[t]
\includegraphics[width=\textwidth]{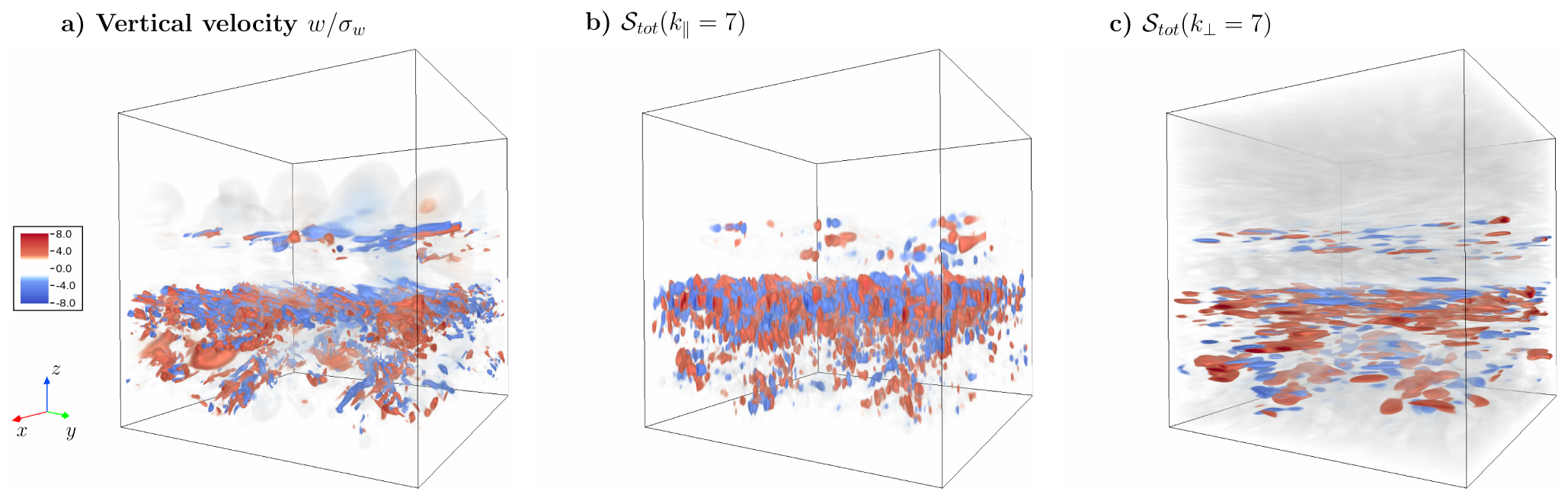}
\caption{Values larger than four standard deviations are highlighted in red (positive) and blue (negative) for the vertical velocity field $w$ (panel a), and the point-wise total energy flux term, vertically filtered at $k_\parallel = 7$, (panel b), and the same term horizontally filtered at the perpendicular wave number $k_\perp$ (panel c). For panels b) and c) positive/negative values mean downscale/upscale energy transfer.
}
\label{fig:ch4_Vz_vs_S_3D}
\end{figure}

In this section we expand the results presented in~\citet{marino_14}, showing how extreme vertical drafts developing in DNS of stratified turbulent flows~\cite{rorai_14,feraco_18} and observed in geophysical flows~\cite{mahrt_89,dasaro_07,chau_21} are able to generate local turbulence, enhancing kinetic and potential energy dissipation. 
Leveraging the {coarse graining} approach, here we assess the feedback of the vertical drafts on the energy transfer and the kinetic-potential energy exchanges locally in the physical space.   
To this end, we consider Run II (see Table~\ref{tab:runs}), whose parameters have been identified in~\cite{feraco_18,Feraco2021} as those associated with the highest level of large-scale intermittency of both velocity and potential temperature fields, assessed through their kurtoses, $K_w$ and $K_\theta$ respectively. A detail of the temporal profile of $K_w$ for Run II is presented in Fig.~\ref{fig:ch4_Kw_run_N8}, in which the red portion of the curve identifies the time interval analyzed in Sec.~\ref{subsec:time_anlysis} and the inset highlights five snapshots of the simulation around a peak of kurtosis used for the following analysis.
Fig.~\ref{fig:ch4_Vz_vs_S_3D} shows three-dimensional renderings of the filtered fields at $t\simeq 178.8\uptau_{NL}$ (see inset in Fig.~\ref{fig:ch4_Kw_run_N8}), a time characterized by a surge of vertical drafts: panel (a) highlights the values of vertical velocity field $w$ larger than four standard deviations ($|w/\sigma_w|>4$),  in red if positive and blue if negative; the total sub-{scale} energy transfer $\mathcal{S}_{tot}$, computed at the cutoff scale $k=7\approx k_B=N/U$ (the latter being the wave number associated with the buoyancy scale) for both parallel and perpendicular integrations of the filter, is presented in panels (b) and (c). Five temporal snapshots between $t=177.9\uptau_{NL}$ and $t=179.7\uptau_{NL}$, have been used to compute the flux terms, corresponding to the red-shaded area in the inset of Fig.~\ref{fig:ch4_Kw_run_N8}. 
Positive values (red), being significantly more numerous and intense for the perpendicular filter, thus as a function of $k_\perp$, indicate a net transfer of the energy to the smaller scales.
Conversely, $\mathcal{S}_{tot}(k_\parallel)$ (panel b) shows almost the same density of structures transferring energy at scales {smaller} (red) and {larger} (blue) than the filtering scale, i.e. $\approx k_B$. Indeed, we will see more quantitatively in the following that around such a scale the net parallel energy transfer almost vanishes $\langle\mathcal{S}_{tot}(k_\parallel\approx k_B)\rangle\sim 0$, that interestingly is related to the typical width of the layers. 
In some regions of the simulation domain, there is a very good correlation between the sub-{scale} term and the extreme values of the vertical velocity. 

\begin{table}[tbh]
\centering
\resizebox{0.4\columnwidth}{!}{%
\begin{tabular}{c|ccccc}
\hline
$|w/\sigma_w|$ & $[0,2.5)$ & $[2.5,3)$ & $[3,4)$ & $[4,6)$ & $[6,\infty)$ \\ \hline
$\%$ volume    & $97.92$   & $0.802$   & $0.704$ & $0.436$ & $0.138$      \\ \hline
\end{tabular}%
}
\caption{Percentage of volume occupied by points having standardized vertical velocity $|w/\sigma_w|$ in a given interval of values. The data are obtained by averaging over the five binaries shown in the inset in Fig.~\ref{fig:ch4_Kw_run_N8}.}
\label{tab:vol_percentage}
\end{table}

\begin{figure*}[t]
\includegraphics[width=\textwidth]{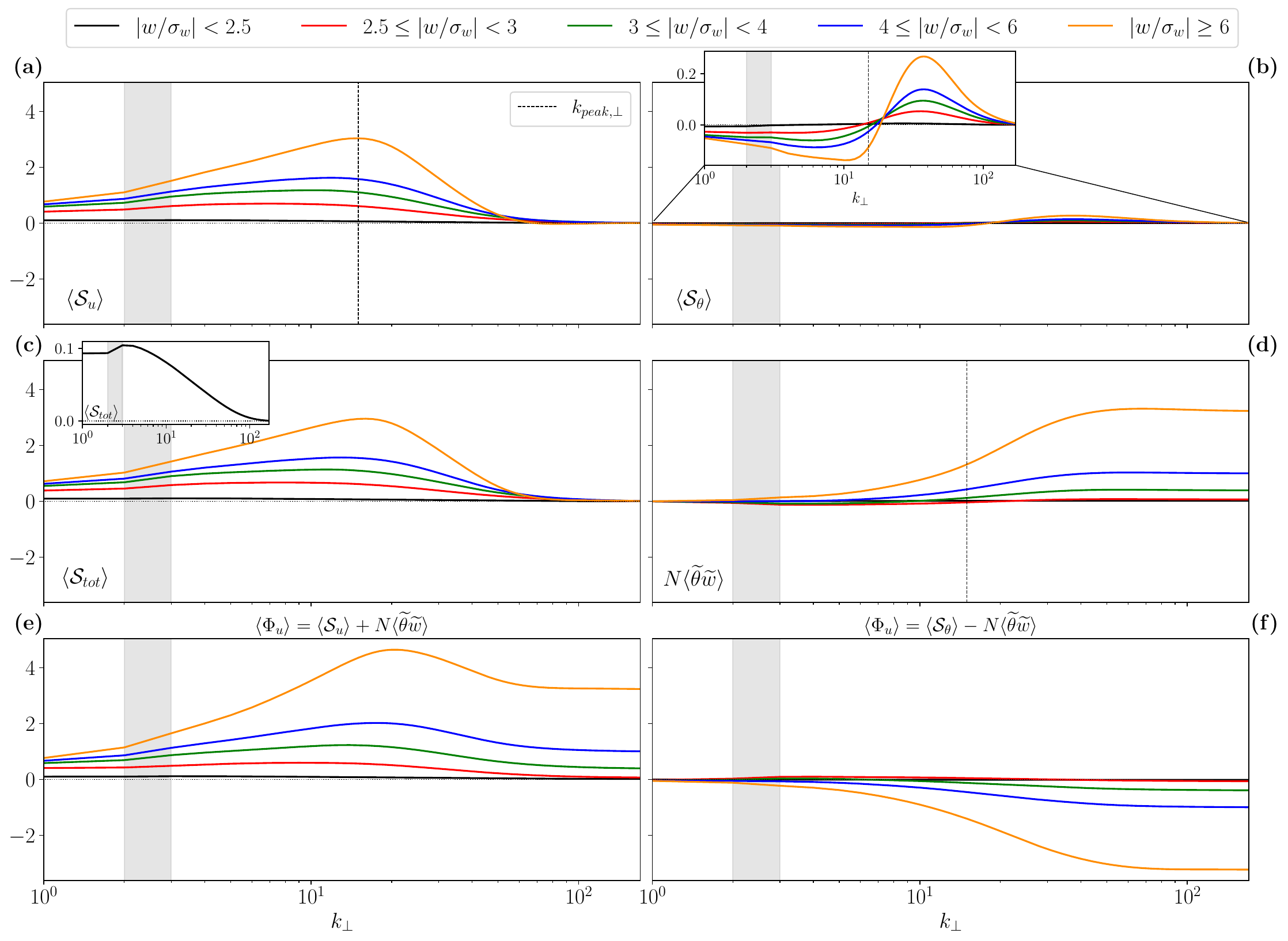}
\caption{Kinetic (a), potential (b), total energy flux terms (c), and buoyancy flux (d) as a function of the filtering wave number $k_\perp=\sqrt{k_x^2+k_y^2}$ for the axisymmetric version of the filtering kernel applied to run II. The gray shaded area indicates the shell where kinetic energy is injected $k_F=[2,3]$ in the simulation through the external forcing. The vertical dashed lines (in panels b, c, and d), at $k_{\rm peak,\perp} = 15 \approx 2k_B$, indicate where the maximal coupling between velocity and potential temperature fields occurs. 
Panels (e) and (f) show the sum of the previous terms for each energy channel, being proportional to the energy flux $\partial_t\langle\widetilde{\mathcal{E}}_{u,\theta}\rangle$ in the inertial range, where the dissipation terms $\langle \mathcal{D}_{\nu,\kappa}\rangle$ are negligible. {The inset in panel (c) emphasizes the total sub-scale term for regions with $|w/\sigma_w|<2.5$.} {Flux terms computed in this sub-domain are weaker and less visible in all the panels.}}
\label{fig:ch4_Sperp_per_drafts_N8}
\end{figure*}

\begin{figure*}[t]
\includegraphics[width=\textwidth]{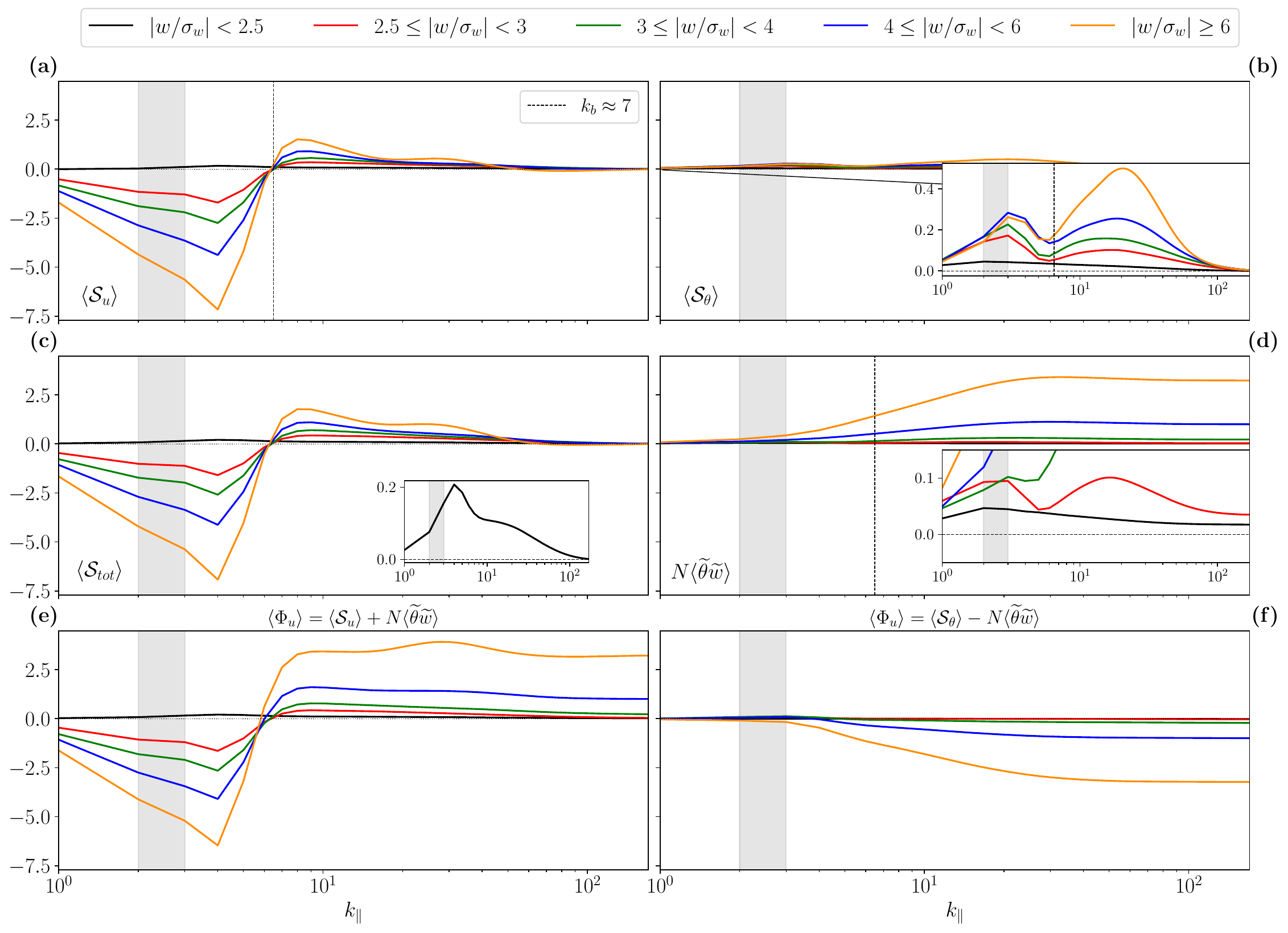}
\caption{Same as Fig.~\ref{fig:ch4_Sperp_per_drafts_N8} for the vertically filtered quantity, here shown as a function of parallel wave numbers $k_\parallel=|k_z|$. The vertical dashed lines (panels b and c) indicate the buoyancy wave number $k_B$. {The  fluxes averaged in the sub-domain identified by the points with $|w/\sigma_w|<2.5$ are in general less intense (see insets in panels b-c).}}
\label{fig:ch4_Spara_per_drafts_N8}
\end{figure*}

In order to quantitatively assess the possibility that the extreme vertical drafts may act as local energy injection mechanism in stratified turbulent flows, we perform averages of all the sub-{scale} terms (i.e. kinetic, potential and buoyancy flux) in sub-domains of the simulation box. In particular, at each time of the simulation the space is partitioned in terms of standardized values of the vertical velocity $|w/\sigma_w|$, then statistical bins are created to accumulate values from $|w/\sigma_w|<2.5$ (corresponding to the domain points with vertical velocities within the Gaussian core of the distribution, accounting for $\sim 97\%$ of the total volume, to $|w/\sigma_w| \in [2.5,3)$, $|w/\sigma_w| \in [3,4)$, $|w/\sigma_w| \in [4,6)$, and finally $|w/\sigma_w| \ge 6$, corresponding to portions of the domain volume characterized by the strongest vertical drafts along the temporal evolution of the simulations. As a reference, values with $|w/\sigma_w|\ge 4$ corresponds to events occurring on average on $\sim 0.6\%$ of the volume under study (see Tab.~\ref{tab:vol_percentage}). 
Number and extension of the vertical velocity bins are constrained by the necessity to have convergent statistics in each bin, at least at the lowest orders (i.e., mean and standard deviation). 

% \ec{La descrizione della figura 4 me la sono persa?}
% \ec{Le spiegazioni dei paragrafi seguenti (A e B) sono parecchio convolute... forse aiuterebbe fare un diagramma tipo figura 1 centrato sul comportamento dei vari S a varie scale ?!}\\
% \rf{L'ho aggiunta a 497--503.}

\subsection{Perpendicular cross-scale energy transfer}
In Fig.~\ref{fig:ch4_Sperp_per_drafts_N8} we report panels showing the nonlinear transfer through scales of the various energy channels averaged over different sub-domains identified by increasing intervals of the standardized vertical velocity $|w/\sigma_w|$ (see legend), and obtained using the perpendicular $k_\perp$ filtering kernel. Panels (a), (b), and (c) show the kinetic $\langle\mathcal{S}_{u}\rangle$, potential $\langle\mathcal{S}_\theta\rangle$ and total energy flux $\langle\mathcal{S}_{\rm tot}\rangle$, respectively, while panel (d) represent the kinetic-to-potential conversion term (or buoyancy flux) $N\langle\wttheta\wtw\rangle$; the two bottom panels show instead the conservative fluxes of kinetic (e) and potential energy (f), proportional in the inertial range to the energy transfer rate, where the dissipation and the large-scale forcing are negligible (see~\eqref{eq:Bssnsq_flow_energy_filtered_averaged} and~\eqref{eq:Bssnsq_potential_energy_filtered_averaged}).
Generally, for all the quantities, higher energy transfer is associated on average with larger values of $|w/\sigma_w|$, at all the scales. The local transfer rate associated with the extreme events (assuming as a reference those are characterized by $|w/\sigma_w|>4$) is on average up to ten times larger than the volume-averaged energy transfer rate for both the kinetic and potential energy channels, i.e. $\langle \varepsilon_{V}\rangle\approx 0.288$ and $\langle \varepsilon_{\theta}\rangle\approx 0.024$.
The cross-scale flux computed over regions with $|w/\sigma_w|<2.5$ (black curve{s} in Fig.~\ref{fig:ch4_Sperp_per_drafts_N8}) shows a peak of transfer close to the forcing shell $k_F \approx 2.5$, as it is expected for a continuously forced simulation {(e.g., see inset in panel c of Fig.~\ref{fig:ch4_Sperp_per_drafts_N8})}.
By looking at the total energy transfer Fig.~\ref{fig:ch4_Sperp_per_drafts_N8} (c) one infers that the forward transfer (i.e., toward smaller scales) is stronger over a wide range of intermediate perpendicular scales, $k_F\lesssim k_\perp\lesssim 40$ ($k_{\rm Oz}\approx 42$ being the Ozmidov scale estimated over the same time interval), in the regions of the domain where the the vertical velocity is higher. The total energy transfer seems also dominated by the kinetic energy contribution, showing the same features and trend (Fig.~\ref{fig:ch4_Sperp_per_drafts_N8}, panel a). There is as well a significant conversion from kinetic to potential energy, approximately over the same range of scales, again  proportionally to values of $|w/\sigma_w|$ (Fig.~\ref{fig:ch4_Sperp_per_drafts_N8}, panel d). At scale $k_\perp \lesssim 10$ the energy conversion is negligible and less sensitive to $|w/\sigma_w|$, while at intermediate scales, $k_B\lesssim k_\perp \lesssim k_{\rm Oz}$, the filtered buoyancy flux $N\langle\wttheta\wtw\rangle$ increases more rapidly the larger the vertical velocity.
Beyond $k_\perp\approx k_{\rm Oz}$, the kinetic-to-potential conversion saturates, meaning that there is no significant exchange between these two energy channels at smaller scales. 
Nevertheless, we mention that $N\langle\wttheta\wtw\rangle$ slightly decreases at $k_\perp\gtrsim k_{\rm Oz}$, though it is difficult to appreciate from the figure, which means that a small portion of the potential energy is converted back into kinetic energy at the smallest scales, especially for high values of $|w/\sigma_{w}|$.
Looking at the potential energy flux term in Fig.~\ref{fig:ch4_Sperp_per_drafts_N8}(b), an interesting phenomenon is observed: a bi-directional potential energy transfer, simultaneously direct (at scales $k_\perp\gtrsim 20$) and inverse (at scales $k_\perp\lesssim 10$) seems to be associated with the emergence of strong vertical velocity draft, strengthening the higher the values of the vertical velocity. 
The wave number at which the sign of the energy transfer switches is close to $k_{\rm peak,\perp} \approx15$ (vertical dashed lines in panels b, c, and d of Fig.~\ref{fig:ch4_Sperp_per_drafts_N8}), the scale at which the energy conversion from kinetic to potential is maximal ($\mathrm{d}\langle\wttheta\wtw\rangle/\mathrm{d}k$), roughly equal to twice the buoyancy wave number ($k_B=N/U\approx 7$). 
This is also the scale at which the peak values of kinetic and total energy transfer are attained (at least in the high vertical velocity bins, $|w/\sigma_w|\geq 4$), as shown in Fig.~\ref{fig:ch4_Sperp_per_drafts_N8} (a,c).
Another length scale that could be associated with the emergence of the bi-directional potential energy flux, representing the maximum vertical distance that can be covered by a fluid parcel before returning to its equilibrium position, is the Ellison scale $\ell_{\rm Ell}=2\pi\theta_{\rm rms}/N$. For run II, $k_{\rm Ell}\approx 17$ which approximately corresponds to the potential flux inversion scale. 
The scenario stemming from this analysis is that -- consistently with the findings in~\citet{marino_22} -- powerful vertical velocity drafts emerging in a parameter space compatible with geophysical flows~\cite{feraco_18} would boost at certain locations of the simulation domain (at intermediate scales) the direct energy transfer already powered by the external large-scale velocity field forcing; on the other hand, through the coupling term $N\wttheta\wtw$, the potential-energy conversion would act as a forcing mechanism to the temperature field (not externally forced), triggering a simultaneous transfer of potential energy to both large and small scales (as shown by $\langle \mathcal{S}_{\theta}\rangle$). This bi-directional transfer is proportional to the intensity of the vertical drafts,  vanishing in the bin corresponding to values $|w/\sigma_w|<2.5$. 

\subsection{Parallel cross-scale energy transfer}
Analogously to what was done in the previous section, here we investigate the energy transfer through the sub-{scale} terms obtained by the application of the filtering kernel depending on the vertical wave numbers $k_\parallel=|k_z|$ only. Even in this case the filter is applied point-wise in the real space, then averages of the energy fluxes are computed over the same sub-domains and time interval. The results are summarized in Fig.~\ref{fig:ch4_Spara_per_drafts_N8}. The behavior of the various sub-{scale} terms is significantly different. 
The different energy channels (i.e., kinetic, potential, and total) in panels (a), (b), and (c), respectively, together with the energy conversion term (panel d), show three distinct regimes. For $k_\parallel\ll k_B$, with $k_B \sim N/U\approx 7$ (vertical dashed lines in panels (b) and (d) of Fig.~\ref{fig:ch4_Spara_per_drafts_N8}), energy is converted from kinetic to potential, on average, for $|w/\sigma_w|<4$, while the opposite happens for regions characterized by the strongest vertical drafts, $|w/\sigma_w|>4$ (see inset in panel d). Indeed, the decreasing trend for $k_\parallel > k_F$ indicates that part of the energy, initially injected by the forcing into the velocity field and converted into potential temperature fluctuations by the buoyancy term, is then being transferred from potential to kinetic in the inertial range until this process saturates at small-scales.
% \ssc{\bf--[anche qui non mi sembra che si veda questo comportamento in $N\langle\wttheta\wtw\rangle$ senza un inset con uno zoom o qualcosa del genere]}
At these scales, for values for $|w/\sigma_w|>2.5$ the total and kinetic energy fluxes (panels a and c) are negative, indicating that energy is transferred to the large scales. A direct transfer -- though much weaker -- is instead detected for values belonging to sub-regions corresponding to the bin $|w/\sigma_w|<2.5$. The wave number at which the direction of the transfer inverts coincides quite accurately with the buoyancy scale $k_B=N/U$.
In the same range a local minimum of the cross-scale potential energy transfer occurs (see inset in panel b of Fig.~\ref{fig:ch4_Spara_per_drafts_N8}), likely related to the fact that part of the potential energy is converted into kinetic energy around this scale. For $k_B<k_\parallel \lesssim k_{\rm Oz}$, on average the energy is converted from kinetic to potential in the entire domain. This range is characterized by a downscale (direct) transfer of kinetic and potential energy, proportional to the magnitude of the vertical velocity, as already observed for $k_\perp$. Finally, for $k_\parallel \gtrsim k_{\rm Oz}$, the direct flux of total and kinetic energy persists, but the conversion is again reversed showing a net but weak flux from potential to kinetic energy. 
The observed behavior of the energy conversion term around the buoyancy scale is in agreement with other numerical studies analyzing Eulerian fields in {spectral} space~\cite{Holloway1988,Staquet1988,Carnevale2001,Brethouwer2007,rorai_14}, as well as Lagrangian velocities and temperatures in the physical space~\cite{Gallon2024}.

\begin{figure*}[t]
\centering
\includegraphics[width=0.9\textwidth]{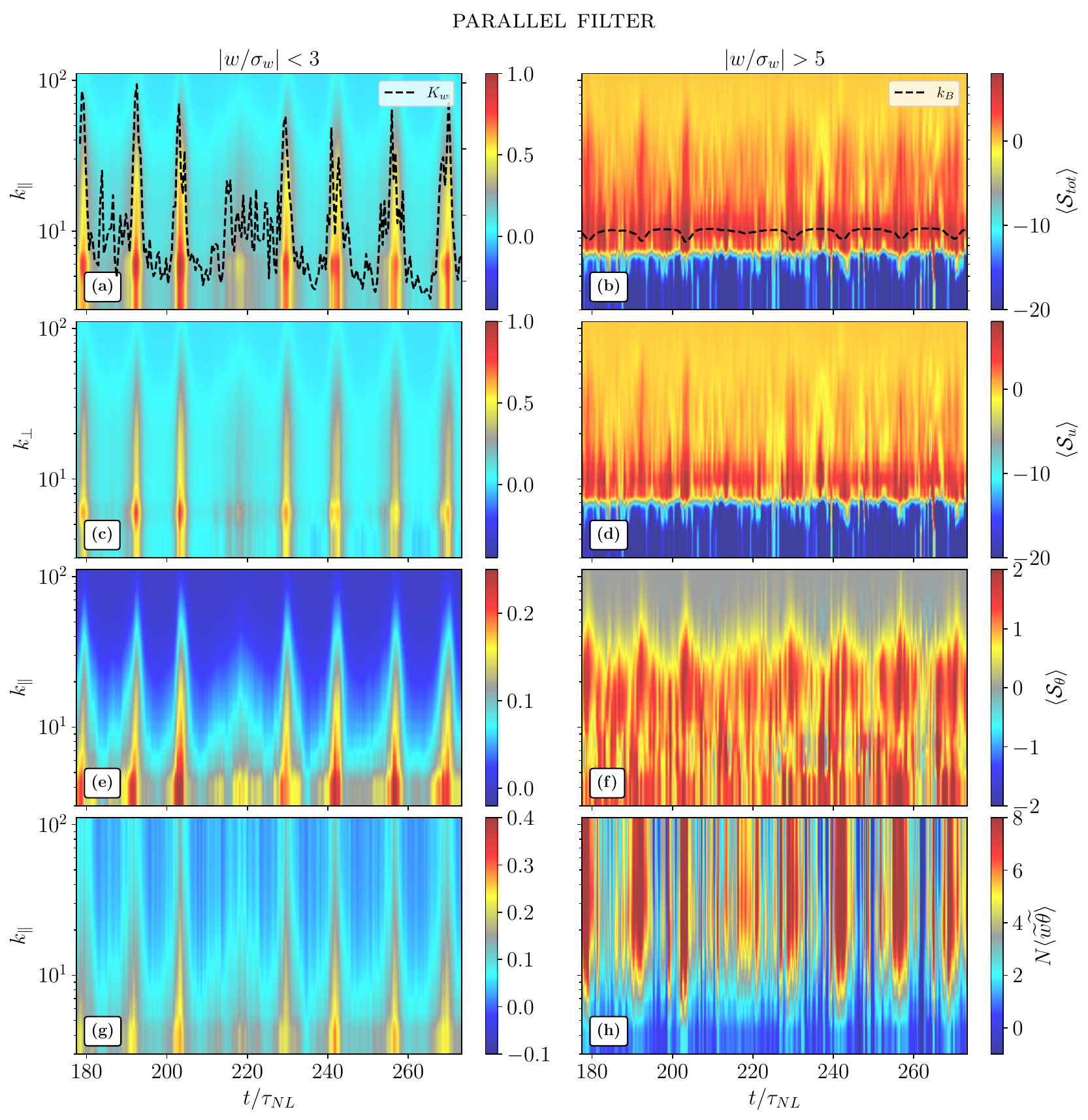}   
\caption{Cross-scale transfer along $k_\parallel=|k_z|$ of total, (a)--(b), kinetic, (c)--(d), and potential energy (e)--(f), as well as the energy conversion term in panels (g)--(h). The left panels show averages computed on regions with $|w/\sigma_w|<3$, the right panels for $|w/\sigma_w|>5$. The black dashed line in panel (a) is the temporal profile of the vertical velocity kurtosis $K_w$, reported for the entire integration time in Fig.~\ref{fig:ch4_Kw_run_N8}, the dashed line in panel (b) is a dissipation-based buoyancy wave number $k_B = N/(\varepsilon_V L)^{1/3}$. Colors in the left panels correspond to values nearly ten times smaller than those in right panels.}
\label{fig:ch4_Spara_vs_t_2D}
\end{figure*}

\subsection{Temporal evolution of the cross-scale energy transfer}\label{subsec:time_anlysis}
In this section more than one hundred turnover times of run II are analyzed, corresponding to the red portion of the vertical velocity kurtosis $K_w$ in Fig.~\ref{fig:ch4_Kw_run_N8}.
The oscillating behavior of $K_w$, with values as high as $K_w\approx 10$ and troughs close to the Gaussian reference (see Fig.~\ref{fig:ch4_Kw_run_N8})
was characterized in~\citet{marino_22} by postulating a fast evolution of the system between two slow manifolds (one associated with waves, the other with the overturning eddy instabilities).
In Fig.~\ref{fig:ch4_Spara_vs_t_2D} we report the temporal variation of the volume-averaged sub-{scale} terms (within the range from $k_\parallel>k_F$ to $k_\parallel\approx k_\eta$) for the two sub-regions corresponding to $|w/\sigma_w|<3$  and $|w/\sigma_w|>5$, left and right panels respectively. For sake of visibility, the palettes of left panels emphasize values nearly ten times smaller than those of the right panels. In panel (a), the temporal variation of $K_w$ appears as a black dashed line
As it was observed for the power spectral density~\cite{marino_22}, which is a second-order quantity, even the energy fluxes -- thus a third-order quantity -- averaged over the entire domain (let us recall that points with $|w/\sigma_w|<3$ represent roughly the 98.7\% of the volume) show a temporal modulation that correlates with the evolution in time of the  kurtosis $K_w$. This further corroborates the evidence presented in~\cite{feraco_18,Feraco2021,marino_22}, that extreme vertical drafts do globally stir the flow, generating local turbulence, enhancing small-scale, dissipation mixing and intermittency. 
By comparing the effect of drafts on regions with $|w/\sigma_w|<3$ and with $|w/\sigma_w|>5$ in Fig.~\ref{fig:ch4_Spara_vs_t_2D}, there is a substantial difference in terms of typical spatial scales, intensity, and overall features. Indeed, for the total and kinetic energy transfers (panels a--d), the regions characterized by vertical drafts are characterized by a bi-directional energy flux, as observed in the previous section, with an inversion scale that is slightly modulated by the intensity of extreme events. The cross-scale potential energy flux is largely positive at any $k_\parallel$ for both regions, with or without extreme events, panels (e) and (f) respectively. However, also in this case substantial differences exist in terms of intensity and scale at which the maximum transfer occurs. For $|w/\sigma_w|<3$ (panel e), the maximum of $\langle \mathcal{S}_\theta\rangle$ always falls at a scale close to the forcing shell $k_F$, as expected, whereas the intensity directly correlates with the vertical velocity kurtosis. On the other hand, in regions with $|w/\sigma_w|>5$ (panel f), the average potential energy transfer rate is pretty constant over the entire time interval, but the scale of the maximum strongly varies with the emergence of extreme events. 
The kinetic-to-potential exchanges mediated by $N\wttheta\wtw$, shown in panels (g) and (h), reflect the persistence of the typical features observed in the previous section. As expected, the most efficient conversion of energy for $|w/\sigma_w|<3$ (panel g) occurs close to the forcing scales; indeed since the runs analyzed are all driven by kinetic energy injection only, the potential temperature fluctuations (null at $t=0$) are energized by the coupling between the two fields, which is maximal at large-scale. The other regions show instead energy conversion peaks in the same range of scales where the the feedback of the vertical drafts is more prominent, i.e. $k_\parallel \gtrsim k_B$., with a strong variation related to $K_w$. This temporal analysis helps further elucidating the role of the velocity in driving the temperature field in the simulations under study, that is enhanced by the emergence of extreme vertical drafts. 

\section{Conclusions}\label{sec:conclusions}
In a certain range of Froude numbers, stratified flows were found to develop in DNS large-scale intermittency, in the form of strong vertical velocity drafts and sudden surges in potential temperature~\cite{rorai_14,feraco_18,Feraco2021}. These events, observed in geophysical flows~\cite{mahrt_89,dasaro_07,lyu_18,chau_21}, are considered extreme as they are characterized by intensities that are several standard deviations larger than their reference average scalar values. Emerging randomly in space and time, generating local turbulence and enhancing dissipation~\cite{marino_22}, they make the flow inhomogeneous, requiring appropriate methodologies to assess their feedback on the energy transfer in the circumscribed regions of the domain where they are detected.    
Here, we employed the {coarse-graining} technique to explore the dynamics of stratified flows characterized by large-scale extreme events. Widely utilized in the literature to analyze neutral~\citep{Aluie_2018,deleo_2022,Buzzicotti_2021} as well as electrically conductive\rf{~\citep{Aluie2010,YangPOF2017,camporeale_18,YangMNRAS2019,Manzini2022a,Manzini2022b,Foldes2024}} turbulent flows, {coarse-graining} approach proved to be a reliable proxy of the classical Fourier flux (which is global-in-space) capable of providing local-in-space information on energy transfers and exchanges across the scales.
In this work, the standard {coarse-graining} procedure was refined and adapted to the Boussinesq framework and its associated energy equations; then, integrations of the flux terms over cylindrical and planar filtering manifolds have been compared with parallel ($k_\parallel=|k_z|$) and perpendicular ($k_\perp=\sqrt{k_x^2+k_y^2}$) energy fluxes computed in the Fourier space, as done for instance in~\citet{marino_14} and~\citet{Alexakis2020}. This test demonstrated the excellent agreement between classical (global) Fourier fluxes and estimates of the energy transfer obtained through the {coarse-graining} by averaging flux terms over the whole domain in stratified flows forced {at intermediate scales {(Sec.\ref{subsec:Fourier_vs_spacefilter})}. 
The {coarse-graining} approach was then implemented on a stably stratified DNS characterized by $Fr\approx0.08$, a value that was shown to be strongly intermittent up to hundreds of turnover times~\cite{Feraco2021,marino_22}, producing the following outcomes:
\begin{enumerate}
    \item The {coarse-graining} analysis revealed that, in regions where powerful vertical velocity drafts develop, enhanced forward kinetic energy transfers are observed at large-intermediate scales -- peaking at scale which is roughly half the buoyancy scale $L_B\sim 1/k_B$ {(see Fig.~\ref{fig:ch4_Sperp_per_drafts_N8})} of the system -- due to the coupling between the ``sub-{scale}'' {turbulent (Reynolds)} stress tensor $\bb{\mathcal{T}}_{uu}$ and the large-scale strain tensor $\boldsymbol{\nabla}\widetilde{\bb{u}}$. 
    % \ssc{\bf--[domanda (magari per un lavoro successivo?): quale dei due termini fa "la maggior parte del lavoro'' in questo trasferimento, $\bb{\mathcal{T}}_{uu}$ o $\boldsymbol{\nabla}\widetilde{\bb{u}}$? o entrambi? e lo stesso pu\'o essere chiesto per i due termini di ${\cal S}_\theta$]}~\rf{Yes! Tra l'altro l'analisi del tensore di Reynolds può essere anche interessante nel contesto delle Large-Eddy Simulation.}
    
    \item In the analyzed simulation, where no external forcing was applied to the (potential) temperature field, vertical velocity drafts act as a mechanism for locally converting energy from kinetic to potential, both along the vertical and horizontal directions in the spectral space; this conversion is mediated by the buoyancy nonlinearity $N\langle \theta w \rangle$ that couples velocity and temperature fields in the Boussinesq framework. Interestingly, this process seems driving a dual transfer of potential energy, simultaneously toward both large and small scales, in the perpendicular direction ($k_\perp$) roughly within the turbulent inertial range. Such a behavior is evidenced by the change of sign of $\langle \mathcal{S}_\theta \rangle$ around $k_\perp \sim 20$, approximately corresponding to the maximum of energy conversion rate from kinetic to potential, as observed through $N\wttheta\wtw$.
    % \ssc{\bf--[sono confuso, adesso stai parlando effettivamente di trasferimento, ma non \'e il termine $N\langle \theta w \rangle$ che fa trasferimento, sarebbe ${\cal S}_\theta$]} 
    These exchanges, quantified here in terms of filtered buoyancy flux, $N\wttheta\wtw$, may also affect the mixing properties of the flow, inducing local variations of the buoyancy flux. 
    
    \item Along the parallel direction in Fourier space ($k_\parallel$), a bi-directional total energy transfer, developing around the buoyancy scale $k_B=N/U\approx 7 $, is associated with the strongest vertical velocity drafts ($|w/\sigma_w|>2.5$). For $k<k_B$ the total energy flux appears indeed to be negative (corresponding to an upscale transfer) and almost twice as intense as the forward energy transfer, occurring at $k>k_B$. These regimes can be explained in terms of different coupling between the velocity and temperature fields. At scales larger than $k_B$, energy conversion associated with regions of strong vertical drafts is, on average, predominantly from potential to kinetic; on the other hand, within the inertial range, the scenario is compatible with what is observed for the perpendicular energy flux. 
    % Differences in the peculiar vertical and horizontal length scales associated with extreme events in the vertical velocity simply reflect the anisotropy of the system.\ssc{\bf--[non ho capito benissimo queste ultime tre frasi... ma potrebbe essere l'orario :) in caso non fosse questo il problema, si pu\'o spiegare meglio?]}~\rf{Non ho capito quali.}
    
\end{enumerate}

The evidence that stratified flows, in a certain region of the parameter space compatible with the atmosphere and the oceans, develop strong large-scale intermittent events in the velocity and temperature fields, being able of mediating energy transfers and conversion (as shown in~\citet{marino_22}), may suggest potential improvements of the parametrization in weather and climate models. Among the physical processes that we plan to include in future extensions of the present study there is certainly rotation, which is critical to describe the dynamics of the Earth's atmosphere at large scales and of the oceans. 
Considering extra terms in the equations would come, of course, with an additional computational cost since, in the presence of forcing, simultaneous direct and inverse energy cascades develop when the Rossby number ($Ro=U_{\rm rms}/\left[f L_{\rm int}\right]$) is small enough~\cite{pouquet_13,marino_15,balwada_2022,Alexakis2024}.
The {coarse-graining} procedure that we proposed here is well-suited for applications to the case of rotating and stratified fluids, as well as to other intermittent, transient, and non-homogeneous turbulent flows observed in nature and in laboratories.

\begin{acknowledgments}
R.F. and R.M. acknowledge support from the ANR project ``EVENTFUL'' (ANR-20-CE30-0011). S.S.C. is supported by the French government through the UCA$^\text{JEDI}$ Investments in the Future project managed by the National Research Agency (ANR) with the reference number ANR-15-IDEX-0001 and by the ANR grant ``MiCRO'' with the reference number ANR-23-CE31-0016. E.C. was partially supported by NASA grants 80NSSC20K1580 ``Ensemble Learning for Accurate and Reliable Uncertainty Quantification" and 80NSSC20K1275 ``Global Evolution and Local Dynamics of the Kinetic Solar Wind". The computing resources utilized in this work were provided by PMCS2I at the \'Ecole Centrale de Lyon. The authors are grateful to Annick Pouquet for the useful discussions during her visit at the \'Ecole Centrale de Lyon, in {M}ay-{J}une 2024. 
\end{acknowledgments}

\bibliography{main.bib}% Produces the bibliography via BibTeX.

\end{document}